\documentclass[twocolumn,english,aps,pra,superscriptaddress,floatfix]{revtex4}
\usepackage[T1]{fontenc}
\usepackage[latin9]{inputenc}
\usepackage{color}
\usepackage{bm}% bold math
\usepackage{xcolor}
\usepackage{amsmath}
\usepackage{amssymb}
\usepackage{graphicx}
\usepackage{natbib}
\usepackage{braket}
\usepackage{psfrag}
\usepackage{tikz}
\usepackage[normalem]{ulem}
\usepackage{qcircuit}
\usepackage[colorlinks=true,urlcolor=blue,citecolor=blue,linkcolor=blue]{hyperref}

  {{\nopagebreak\hspace*{\fill}$\Box$\par\vspace{12pt}}}

\newcommand{\be}{\begin{equation}}
\newcommand{\ee}{\end{equation}}

\newcommand\blfootnote[1]{%
  \begingroup
  \renewcommand\thefootnote{}\footnote{#1}%
  \addtocounter{footnote}{-1}%
  \endgroup
}

\begin{document}

%\preprint{}

\title{Quantum Algorithms for Mixed Binary Optimization\\
applied to Transaction Settlement
\blfootnote{
\textcopyright ~Barclays and International Business Machines Corporation 2019.
BARCLAYS is a registered trade mark of Barclays plc, all rights are reserved.
IBM and IBM Q are trademarks of International Business Machines Corporation, registered in many jurisdictions worldwide.  Other product or service names may be trademarks or service marks of Barclays, IBM or other companies.
}
}% Force line breaks with \\
%\thanks{}%

\author{Lee Braine}
\affiliation{%
Barclays
}%
\author{Daniel J. Egger}
\affiliation{%
IBM Research -- Zurich
}%
\author{Jennifer Glick}
\affiliation{%
IBM T.J.~Watson Research Center
}%
\author{Stefan Woerner}
\email{wor@zurich.ibm.com}
\affiliation{%
IBM Research -- Zurich
}%

\date{\today}% It is always \today, today,
             %  but any date may be explicitly specified

\begin{abstract}
We extend variational quantum optimization algorithms for Quadratic Unconstrained Binary Optimization problems to the class of Mixed Binary Optimization problems.
This allows us to combine binary decision variables with continuous decision variables, which, for instance, enables the modeling of inequality constraints via slack variables.
We propose two heuristics and introduce the Transaction Settlement problem to demonstrate them.
Transaction Settlement is defined as the exchange of securities and cash between parties and is crucial to financial market infrastructure.
We test our algorithms using classical simulation as well as real quantum devices provided by the IBM Quantum Computation Center.
\end{abstract}

%\pacs{}% PACS, the Physics and Astronomy
				% Classification Scheme.
%\keywords{Suggested keywords}%Use showkeys class option if keyword
                              %display desired
\maketitle

%\tableofcontents

%%%%%%%%%%%%%%%%%%%%%%%%%%%%%%%%%%%%%%%%%%%%%%%%%%%%%%%%%%%%%%%%%%%%%%%%%%%
\section{\label{sec:introduction} Introduction}
%%%%%%%%%%%%%%%%%%%%%%%%%%%%%%%%%%%%%%%%%%%%%%%%%%%%%%%%%%%%%%%%%%%%%%%%%%%

Quantum computers process information using the laws of quantum mechanics.
They are well suited for a number of tasks such as simulating quantum mechanical systems \cite{kandala2017hardware, Ganzhorn2019} and factoring \cite{Shor1997}.
Additionally, quantum computers can provide a quadratic speed-up over classical Monte-Carlo simulations which may be used to evaluate risk \cite{Woerner2018, Egger2019} and price financial derivatives \cite{Stamatopoulos2019, Rebentrost2018}.
Another possible application area for quantum computers is optimization, particularly combinatorial optimization.
It is not believed that quantum computers will be able to solve NP-hard problems in polynomial time \cite{bernstein1997quantum}.
There is, however, a significant effort in designing quantum heuristics that could be practically useful by finding near optimal solutions \cite{Peruzzo2014, Farhi2014, Nannicini2019, Barkoutsos2019}.
Algorithms such as the \emph{Variational Quantum Eigensolver} (VQE) \cite{Peruzzo2014} and the \emph{Quantum Approximate Optimization Algorithm} (QAOA) \cite{Farhi2014} are designed to tackle \emph{Quadratic Unconstrained Binary Optimization} (QUBO) problems.
However, many relevant problems in business and science are \emph{Mixed Binary Optimization} (MBO) problems, with discrete and continuous variables, or with constraints that cannot be modelled as part of a QUBO problem, e.g.~inequality constraints.
In this paper, we introduce an approach to extend the existing quantum methods to more general MBO problem classes.

We test our algorithm on the \emph{Transaction Settlement} problem by focusing on securities settlement in capital markets.
Here, transaction settlement is the process in which securities (tradeable financial assets such as shares, bonds and derivatives) are delivered usually against a payment.
This exchange between parties can be facilitated by a clearing house, which also mitigates the counterparty risk \cite{Loader2002}.
Financial institutions submit the details of the trades (e.g. buy $x$ shares of some company for an amount $y$ of some currency) to the clearing house, which runs a complex optimization algorithm on the resulting batch of transactions while taking into account credit and collateral facilities.
The objective is typically to settle as many transactions as possible or to maximize the total value of the settled transactions.
Transaction settlement is a difficult optimization problem due to a combination of both the legal constraints that must be satisfied when settling delivery-versus-payment transactions and the additional optionality introduced by collateralizing assets and utiltizing credit facilities. A variety of approaches are currently employed, ranging in complexity from basic gross settlement systems (which settle on a simple transaction-by-transaction basis) to sophisticated probabilistic techniques such as simulated annealing (whereby an optimization process is performed to identify a sufficient subset of transactions that can settle and then that subset is actually settled). This is an industry process of systemic importance because of the volume and value of transactions settled, e.g.~over \$1.85 quadrillion of securities transactions were processed in 2018 by subsidiaries of the post-trade market infrastructure DTCC \cite{dtcc2018}.

Our manuscript is structured as follows.
In Sec.~\ref{sec:mbo}, we define QUBO problems and discuss their relation to MBO problems.
In Sec.~\ref{sec:algorithms}, we introduce hybrid quantum-classical optimization algorithms for MBO by extending the known algorithms.
In Sec.~\ref{sec:transaction_settlement} we map the transaction settlement problem to this framework and, in Sec.~\ref{sec:results}, we demonstrate the performance of our hybrid algorithm on concrete instances using cloud-based quantum hardware.
We conclude in Sec.~\ref{sec:conclusions} and discuss open questions and directions of further research.

%%%%%%%%%%%%%%%%%%%%%%%%%%%%%%%%%%%%%%%%%%%%%%%%%%%%%%%%%%%%%%%%%%%%%%%%%%%
\section{\label{sec:mbo} Mixed Binary Optimization}
%%%%%%%%%%%%%%%%%%%%%%%%%%%%%%%%%%%%%%%%%%%%%%%%%%%%%%%%%%%%%%%%%%%%%%%%%%%

QUBO problems, defined as
\begin{eqnarray}
    \min_{x \in\{0, 1\}^n} && x^T A x + b^T x + c, \label{eq:qubo}
\end{eqnarray}
where $A \in \mathbb{R}^{n \times n}$, $b \in \mathbb{R}^n$, and $c \in \mathbb{R}$, have a wide range of applications such as portfolio optimization or the traveling salesman problem, but are hard to solve \cite{lucas2014ising, Barkoutsos2019}.
Variational quantum algorithms for combinatorial optimization can find good solutions to QUBO problems once mapped to an Ising Hamiltonian $H$ by setting $x_i = (1 - z_i)/2$ for $z_i \in \{-1, +1\}$ and replacing $z_i$ by Pauli Z-matrices $\sigma_i^Z$.
This allows us, for instance, to use VQE or QAOA to approximate the ground state of $H$, which corresponds to the optimal solution of the QUBO problem.

Since many industry relevant problems, such as portfolio optimization with budget constraints \cite{Venturelli2018a} or the knapsack problem and its variants \cite{Kellerer2004}, are not QUBO problems, we also consider MBO problems, here defined as
\begin{eqnarray}
\min_{\substack{x \in \{0, 1\}^n \\ y \in Y}} && x^T A(y) x + b(y)^T x + c(y), \label{eq:mbo}
\end{eqnarray}
where $Y \subset \mathbb{R}^m$ is the feasible set for the continuous variables $y$ and $A: \mathbb{R}^m \rightarrow \mathbb{R}^{n \times n}$, $b: \mathbb{R}^m \rightarrow \mathbb{R}^{n}$ and $c: \mathbb{R}^m \rightarrow \mathbb{R}$ are given functions of $y$.
MBO problems therefore combine discrete and continuous variables $x$ and $y$, respectively. Furthermore, for a fixed $y \in Y$, the MBO problem~(\ref{eq:mbo}) corresponds to the QUBO problem~(\ref{eq:qubo}).
While other definitions of MBO exist, we have chosen this one to be compatible with the variational heuristics we want to extend.

%Existing variational quantum algorithms for combinatorial optimization can be applied to find good solutions to Quadratic Unconstrained Binary Optimization (QUBO) problems \WOR{references}:
%\begin{equation}
%    \min_{x \in\{0, 1\}^n} x^T A x + b^T x, \label{eq:qubo}
%\end{equation}
%where $b \in \mathbb{R}^n$ and $A \in \mathbb{R}^{n \times n}$.
%Setting $x_i = (1 - z_i)/2$ for $z_i \in \{-1, +1\}$ and replacing $z_i$ by Pauli Z-matrices $\sigma_i^Z$ allows to map a QUBO to an Ising Hamiltonian $H$ and then use, for instance, the Variational Quantum Eigensolver (VQE) \cite{Peruzzo2014} or the Quantum Approximate Optimization Algorithm (QAOA) \cite{Farhi2014} to approximating the ground state of $H$, which can be translated to a solution of the corresponding QUBO.

%Although QUBOs represent a very interesting problem class with relevant applications such as \WOR{examples + references}, there are many other important problems that are not accessible.
%Particularly, problems that include both, binary and continuous variables..

%Therefore, we extend QUBO to Mixed Binary Optimization (MBO):
%\begin{eqnarray}
%\min_{\begin{array}{c} x \in \{0, 1\}^n \\ y \in Y\end{array}} && x^T A(y) x + b(y)^T x + c(y), \label{eq:mbo}
%\end{eqnarray}
%where $Y \subset \mathbb{R}^m$ is assumed to be the feasible set for the continuous variables and $A: \mathbb{R}^m \rightarrow \mathbb{R}^{n \times n}$, $b: \mathbb{R}^m \rightarrow \mathbb{R}^{n}$ and $c: \mathbb{R}^m \rightarrow \mathbb{R}$ are given functions.
%In other words, for a fixed $y \in Y$, Problem (\ref{eq:mbo}) corresponds to a QUBO.

In QUBO problems we model linear equality constraints by adding quadratic penalty terms of the form $(u^T x + v)^2$, with given $u\in \mathbb{R}^{n}$ and $v\in \mathbb{R}$, scaled by a large weight $\lambda$, to the objective. 
This forces optimal solutions to satisfy $u^T x + v = 0$.
Other types of constraints are usually not possible in QUBO, since they cannot be represented as quadratic penalty terms.
In the \emph{Quantum Alternating Operator Ansatz} the mixing operator in QAOA is adjusted to keep the process in the feasible set assuming a feasible initial state \cite{Hadfield2017}. 
This is a promising, but non-generic, approach to incorporate constraints.

Since MBO problems allow continuous variables, we may introduce a slack variable $s \geq0$ to enforce the equality constraint $u^T x + v + s = 0$. 
This is equivalent to the inequality constraint $u^T x + v \leq 0$, without $s$.
Thus, by allowing binary and continuous variables we extend the type of constraints we can model.
The lower bound on $s$ needs to be included as a continuous constraint defining the feasible set $Y$.

%\sout{It would be possible to generalize the definition of MBO problems even further. However, we refrain from doing so due to the algorithms we introduce and discuss in the following section.}

%Although the MBO problem in (\ref{eq:mbo}) can be generalized, we only consider MBO problems of the form (\ref{eq:mbo}) to be able to use the variational quantum algorithms.

%%%%%%%%%%%%%%%%%%%%%%%%%%%%%%%%%%%%%%%%%%%%%%%%%%%%%%%%%%%%%%%%%%%%%%%%%%%
\section{\label{sec:algorithms} Quantum Optimization Algorithms for MBO}
%%%%%%%%%%%%%%%%%%%%%%%%%%%%%%%%%%%%%%%%%%%%%%%%%%%%%%%%%%%%%%%%%%%%%%%%%%%

Variational quantum algorithms for QUBO first translate the problem into an Ising Hamiltonian $H$, as discussed in Sec.~\ref{sec:mbo}.
Classical optimization can then find the values of the parameters $\theta$ of a trial wavefunction $\ket{\psi(\theta)}$ to, for instance, minimize the expected value
\begin{eqnarray}
    \min_{\theta} && \bra{\psi(\theta)} H \ket{\psi(\theta)}.
\end{eqnarray}
The expected value is easily estimated by repeatedly measuring $\ket{\psi(\theta)}$ along the Z-axis as $H$ only consists of $\sigma^Z$-terms.
Since we consider classical optimization, we can translate every measurement of $\ket{\psi(\theta)}$ into a $n$-bit string and directly evaluate (\ref{eq:qubo}).
If we measure $\ket{\psi(\theta)}$ $N$-times and denote the resulting bit strings by $x^j(\theta)$ where $j=1, \ldots, N$, the resulting optimization problem is 
\begin{eqnarray}\label{eq:avg_vqe_qubo}
    \min_{\theta} && \sum_{j=1}^N x^j(\theta)^T A x^j(\theta) + b^T x^j(\theta) + c.
\end{eqnarray}
We may improve the performance of the optimization algorithm by replacing the sample average in problem (\ref{eq:avg_vqe_qubo}) by another aggregation function such as the Conditional Value at Risk (CVaR). That is, for a given $\alpha \in (0, 1]$, we only average over the best $\alpha N$ samples, which can help the classical optimization process find better results \cite{Barkoutsos2019}.

%Variational quantum algorithms for QUBOs first translate the problem into an Ising Hamiltonian $H$, define trial wavefunctions $\ket{\psi(\theta)}$, and then use classical optimization to find good parameters $\theta$ to minimize for instance the expected value, i.e.,
%\begin{eqnarray}
%    \min_{\theta} \bra{\psi(\theta)} H \ket{\psi(\theta)}.
%\end{eqnarray}

In this paper, we consider two variational algorithms. First, VQE with a generic trial wavefunction $\ket{\psi(\theta)}$ and second, QAOA, where  we construct $\ket{\psi(\theta)}$ based on $H$.
In the following, we discuss how to extend these algorithms to be applicable to MBO problems.

If a generic trial-solution $\ket{\psi(\theta)}$ is given, as in VQE, we can map a MBO problem to a continuous optimization problem
\begin{eqnarray}
\min_{\substack{ \theta \\ y \in Y}} && \sum_{j=1}^N x^j(\theta)^T A(y) x^j(\theta) + b(y)^T x^j(\theta) + c(y) \nonumber \\
\label{eq:mbo_vqe_sample_average}
\end{eqnarray}
for which solutions can be approximated using classical optimization schemes.

Let us now consider a rule to derive a trial wavefunction $\ket{\psi_{H}(\theta)}$ from a given Hamiltonian $H$ such as given by QAOA.
For MBO, every fixed $y \in Y$ defines a QUBO and can, thus, be translated to an Ising Hamiltonian $H(y)$.
This allows us to define trial wavefunctions $\ket{\psi_{H(y)}(\theta)}$ that now depend on $y$.
As in (\ref{eq:mbo_vqe_sample_average}) we can translate the MBO to a continuous optimization problem, where the measurements $x^j(\theta, y)$ depend on $\theta$ as well as $y$.
In summary, $A(y)$, $b(y)$, and $c(y)$ define the Hamiltonian $H(y)$, which in turn defines the $\theta$- and $y$-dependent trial wavefunction $\ket{\psi_{H(y)}(\theta)}$.
These two approaches allow us to extend the existing quantum heuristics from QUBO to MBO enabling us to test variational quantum algorithms on a larger problem class. 

Next, we introduce a heuristic that is designed to handle slack variables resulting from modelling inequality constraints explicitly.
This can help the classical optimizer used to solve \eqref{eq:mbo_vqe_sample_average} to move out of local minima.
Suppose, for simplicity, that we have a problem with only binary variables and a single slack variable, coming from an inequality constraint, formally given by
\begin{eqnarray}\label{eqn:lambda}
    \max_{\substack{ x \\ s \geq 0}} x^T A x + b^T x + c + \lambda \left(u^T x + v + s \right)^2.
\end{eqnarray}
The large scalar $\lambda > 0$ enforces $u^T x + v + s = 0$.
The new heuristic consists of repeating the following three steps a fixed number of times, or until some pre-defined termination criterion is met:
\begin{enumerate}

    \item Perform a preset fixed number of iterations of a classical optimizer to jointly optimize $\theta$ and $s$, and store the resulting samples $x^j(\theta)$ from the last iteration.
    
    \item Analyze every individual sample $x^j(\theta)$ and define
    \begin{eqnarray}
        s^j = \max \left\{0, -(u^T x^j(\theta) + v) \right\}.
    \end{eqnarray}
    In other words, for each sampled binary vector $x^j(\theta)$, we derive the slack variable $s^j$ such that the constraint $u^T x + v + s = 0$ is satisfied, if possible, or such that we minimize the violation.
    Evaluate the overall objective value for $x^j(\theta)$ and $s^j$ and identify the candidate solution that performs best, i.e., that achieves the smallest objective value.
    
    \item Fix the $s^j$ determined in the previous step and rerun the classical optimizer over $\theta$ only for a preset fixed number of iterations.

\end{enumerate}

This simple heuristic can help significantly to move out of local optima of the considered continuous optimization problem in $\theta$ and $s$.
The algorithmic parameters, such as number of cycles and number of iterations per cycle need to be determined according to the problem of interest.
As we show in the following section, applying a classical optimizer directly to solve (\ref{eq:mbo_vqe_sample_average}) often gets stuck in local optima, while our new heuristic reliably found the globally optimal solution of the considered test case.
The introduced algorithm can be easily extended to more complex problems with additional continuous variables and multiple inequality constraints/slack variables.

%%%%%%%%%%%%%%%%%%%%%%%%%%%%%%%%%%%%%%%%%%%%%%%%%%%%%%%%%%%%%%%%%%%%%%%%%%%
\section{\label{sec:transaction_settlement} Transaction Settlement}
%%%%%%%%%%%%%%%%%%%%%%%%%%%%%%%%%%%%%%%%%%%%%%%%%%%%%%%%%%%%%%%%%%%%%%%%%%%

\begin{figure*}[t]
\centering
\includegraphics[width=\linewidth]{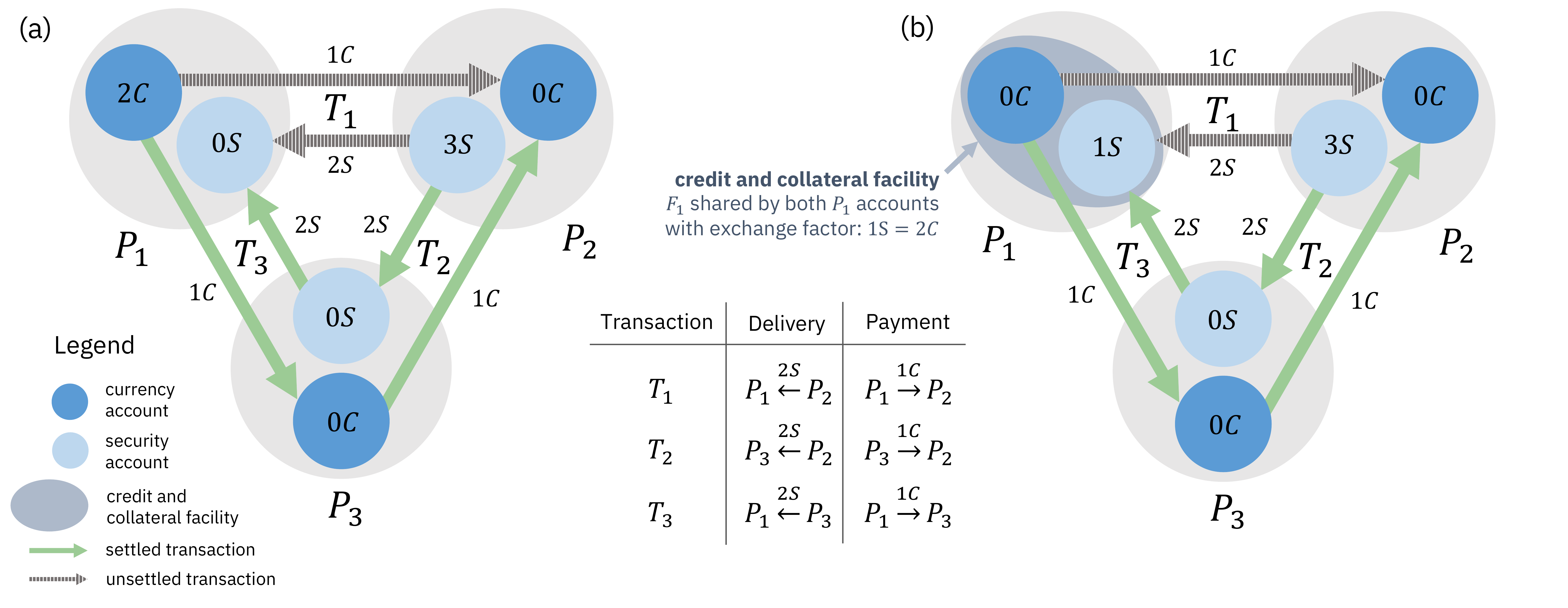}
\caption{Diagrams illustrating a batch of three delivery-versus-payment transactions $T_1, T_2, T_3$ between three parties $P_1, P_2, P_3$. Each party has a currency account $C$ and security account $S$, with amounts shown in the interior circles. Each transaction is shown with a pair of arrows and the corresponding amount of security and currency involved. In (b) we consider a simplified credit and collateral facility (with a net zero amount of credit) shared by both accounts of $P_1$, with an exchange factor between currency and security of $1 S = 2 C$. In both (a) and (b), the optimal solution settles transactions $T_2$ and $T_3$, as indicated by the pairs of solid green arrows. With the modification of $P_1$'s account balances in (b), none of the transactions could settle without the presence of the credit and collateral facility $F_1$, which $P_1$ can use to convert security into currency to facilitate the transaction with $P_3$.}
\label{fig:settlement-1-2-diagrams}
\end{figure*}

In this section, we introduce the Transaction Settlement problem in which parties continuously submit the details of the trades to a clearing house. 
All of the corresponding delivery-versus-payment transactions recorded by the clearing house during a predefined time interval, plus all the unsettled transactions remaining from the previous batch, form a new batch of $I$ transactions submitted by $K$ parties.
$\Gamma_k$ is the set of transactions of party $k$. The clearing house must then settle as many transactions as possible within the given batch. The $i$th transaction involving a party $k$ is described by a sparse vector $\vec{v}_{ik}$ of length $J$, the total number of security and currency accounts, with only two nonzero entries, one for the security (delivery) and one for the currency (payment). The transaction vectors of two parties $k$ and $k'$ that participate in a transaction $i$ sum to zero: $\vec{v}_{ik} + \vec{v}_{ik'} = \vec{0}$. To each transaction $i$ we associate a binary variable $x_i$ to indicate settlement.

\begin{figure*}[t]
\centering
\includegraphics[width=0.75\linewidth]{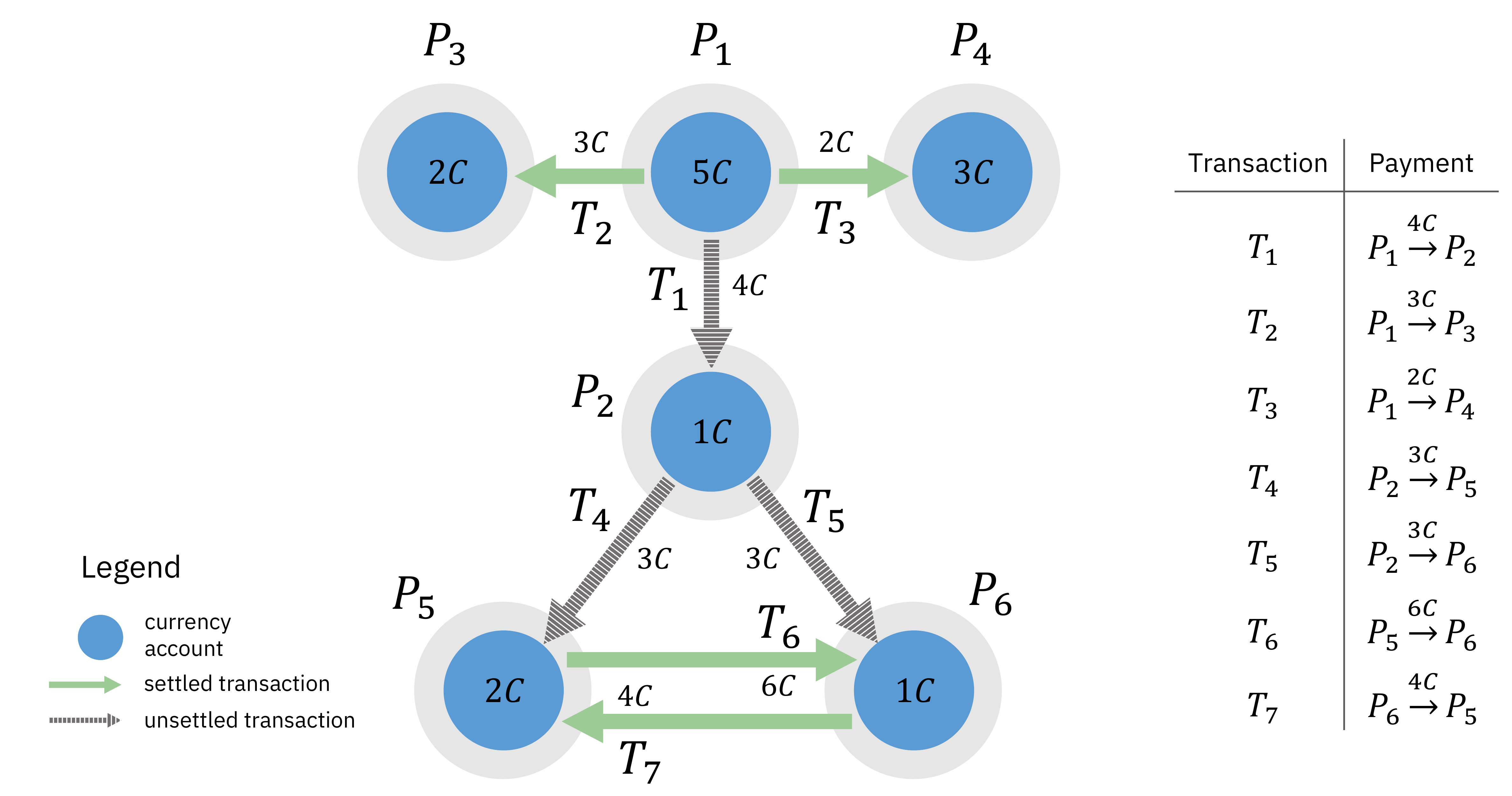}
\caption{Diagram illustrating a batch of seven payments $T_1, \ldots, T_7$ between six parties $P_1, \ldots, P_6$. Each party has a single currency account $C$, with amounts shown in the interior circles. Each transaction is shown with a single arrow and the corresponding amount of currency involved. Several optimal solutions exist that settle four total transactions, e.g., payments $T_2, T_3, T_6, T_7$, shown here with solid green arrows to differentiate the unsettled transactions indicated by grey hatched arrows.}
\label{fig:settlement-3-diagrams}
\end{figure*}

We, thus, seek to maximize the weighted sum of settled transactions
%The optimization problem to determine which trades settle can be written as follows:
\begin{eqnarray} \label{eq:transaction_settlement_1_ineq}
\max_{\vec{x}} && \sum_{i=1}^I w_i x_i \\
\textrm{subject to:} && \vec{b}_k + \sum_{i \in \Gamma_k} x_i \vec{v}_{ik} \geq \vec{l}_k,~ \forall_k. \nonumber
\end{eqnarray}
The weight $\omega_i$ may, for instance, be the total monetary value of transaction $i$, or $\omega_i=1~\forall i$ if we seek to optimize the number of settled transactions.
The $K$ vector constraints in \eqref{eq:transaction_settlement_1_ineq} feature the balance vector $\vec{b}_k$, which encodes the securities and currencies that party $k$ owns before any transactions settle.
Additionally, party $k$ may be granted credit for some or all accounts, which is encoded by $\vec{l}_k$.
There are $K\times J$ constraints in \eqref{eq:transaction_settlement_1_ineq}, but, in practice, the sparsity of $\vec{b}_k$ $\vec{v}_{ik}$, and $\vec{l}_k$ reduces the number of constraints.
Thus, the optimization described by Eq.~\eqref{eq:transaction_settlement_1_ineq} must ensure that the credit limits are not exceeded.
The transaction settlement optimization problem is complex and may feature dependency chains and even cyclical dependencies between transactions, as seen in Figs.~\ref{fig:settlement-1-2-diagrams} and~\ref{fig:settlement-3-diagrams}.

\begin{center}
\begin{table}[t]
\begin{tabular}{ l  c }
Inputs &  \\ \hline
transactions & $i \in \{1, \ldots, I\}$ \\
parties & $k \in \{1,\ldots, K\}$ \\
transactions of party $k$ & $\Gamma_k$ \\
currencies/securities & $j \in \{1, \ldots, J\}$ \\
transaction values & $\vec{v}_{ik} \in \mathbb{R}^J$ \\
transaction weights & $\vec{w} \in \mathbb{R}^I_{\ge0}$ \\
credit limits & $\vec{l}_k \in \mathbb{R}^J$ \\
balance & $\vec{b}_k \in \mathbb{R}^J$ \\
exchange factor & $\vec{r}_k \in \mathbb{R}^J_{\ge0}$\\
credit pool limit & $p_k \in \mathbb{R}_{\le 0}$\\\\

Decision Variables & \\ \hline
settle transaction or not & $\vec{x} \in \{0,1\}^I$\\
slack variables & $\vec{s}_k \in \mathbb{R}^J_{\ge0}$\\
\end{tabular}
\caption{The notation used throughout this paper to define the Transaction Settlement optimization problem.}
\label{tab:notation}
\end{table}
\end{center}

Each account (associated with a security or currency) owned by a party can optionally participate in a simplified credit and collateral facility, which permits the account to contribute to a collateral pool and also use the resulting credit pool. Several accounts owned by a party can participate in the same credit and collateral facility, permitting shared pools (in principle, several credit and collateral facilities can be defined for each party). The accounts owned by party $k$ may add to or draw from the credit pool such that a total credit amount $p_k$ is not exceeded. This can be modeled as an additional continuous constraint
% If we take credit limits $\vec{l}_k$ to be negative, then also take $p_k$ negative, and use $\ge$.}:
\begin{equation}
 \label{eqn:pool_constraint}
 \vec{r}_k \cdot \vec{l}_k \ge p_k,
\end{equation}
for each party $k$ that participates in a credit and collateral facility. The exchange factors $\vec{r}_k$ represent the conversion of a currency/security in one account into a currency/security of another account when exchanged via the facility and may be party dependent. In this scenario, the credit limits $\vec{l}_k$ can now be variable, instead of fixed for each account, to reflect the amount contributed to or drawn from the shared pool. 
The notation we use to formulate this problem is summarized in Tab.~\ref{tab:notation}.

We now transform problem \eqref{eq:transaction_settlement_1_ineq}, following Sec.~\ref{sec:mbo}, into the form given in (\ref{eq:mbo}) which may be solved using the variational quantum solvers discussed in Sec.~\ref{sec:algorithms}.
Non-negative slack variables $\vec{s}_k$ are used to cast the inequality constraints into equality constraints, i.e. \eqref{eq:transaction_settlement_1_ineq} becomes
\begin{eqnarray} \label{eq:transaction_settlement_1}
\max_{\substack{\vec{x} \\ \vec{s}_k \geq 0}} && \sum_{i=1}^I w_i x_i \\
\textrm{subject to:} && \vec{b}_k + \sum_{i \in \Gamma_k} x_i \vec{v}_{ik} = \vec{l}_k + \vec{s}_k,~ \forall_k. \nonumber
\end{eqnarray}

Next, we map this constrained optimization problem to an unconstrained optimization problem by transforming the equality constraint into a quadratic penalty term scaled by a large number $\lambda \ge 0$:
\begin{equation}\label{eq:transaction_settlement_2}
\max_{\substack{\vec{x} \\ \vec{s}_k \geq 0}} \left[ \sum_{i=1}^I w_i x_i - \lambda \sum_{k=1}^K \left( \vec{b}_k + \sum_{i \in \Gamma_k} x_i \vec{v}_i - \vec{l}_k - \vec{s}_k \right)^2 \right].
\end{equation}
In this form, we can solve the optimization problem using the methods of Sec.~\ref{sec:algorithms}, where the number of qubits needed to represent the trial wavefunction is equal to the number of transactions under consideration. The non-negativity constraint on the slack variables as well as constraint \eqref{eqn:pool_constraint} are handled by the classical optimizer as they do not involve the decision variables $x_i$.
Our formulation of the transaction settlement problem assumes that the order in which transactions are settled within a batch can be ignored. 
Therefore, only the settled transactions affect the balances.

We now describe the Transaction Settlement problem for three concrete test cases, with numerical results discussed later in Sec.~\ref{sec:results}. The first two test cases involve three transactions between three parties of a single currency and security, while the third test case is a batch of seven payments of a single currency between six parties. In all three problems, we maximize the total \emph{number} of transactions. 

\begin{figure*}[ht]
\centering
\includegraphics[width=0.85\linewidth]{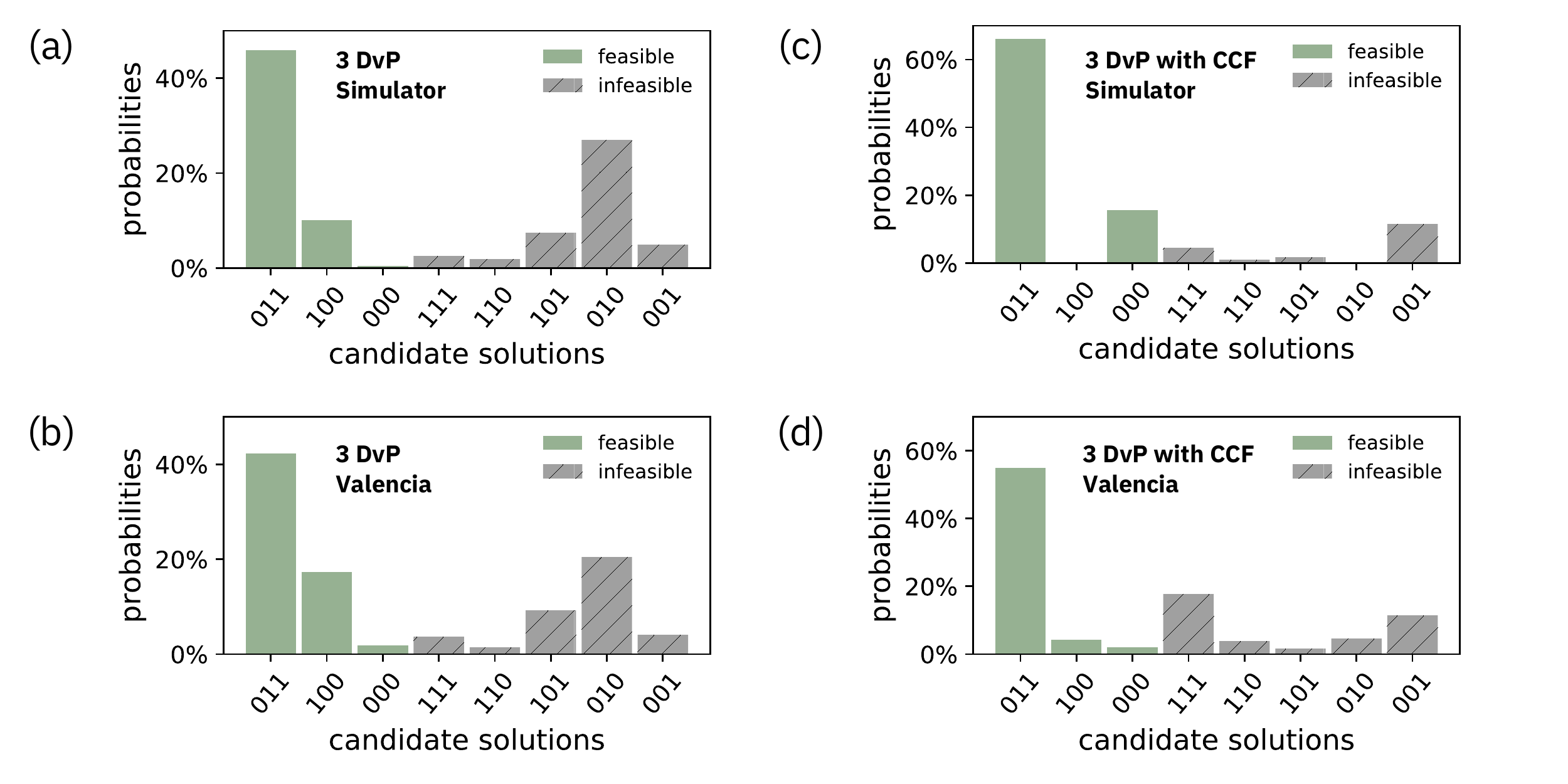}
\caption{Results for (a, b) the three delivery-versus-payment (DvP) transactions of Eq.~\eqref{eq:settlement-1-objective} and (c, d) the three DvP transactions with a credit and collateral facility (CCF) of Eq.~\eqref{eq:settlement-2-objective}. (a, c) show the results from classical simulation, while (b, d) show the results from the \emph{Valencia} 5-qubit quantum processor. Candidate solutions are grouped into feasible (green bars) and infeasible (grey hatched bars) sets and then sorted left to right by descending value of the objective function (number of trades settled).
In all four cases, the optimal solution $(x_1, x_2, x_3) = 011$, indicating that transactions $T_2$ and $T_3$ settle, is sampled with the highest probability. 
Both problems used the $R_y$ variational circuit of depth $d=2$ shown in Fig.~\ref{fig:circuit}, a penalty factor $\lambda=10^3$ for adding constraints to the objective function, the CVaR aggregation function with $\alpha = 25\%$, and 150 iterations of COBYLA (without additional cycles to explicitly handle slack variables, see Sec.~\ref{sec:algorithms}).
%\JRG{[numerics details: statevector simulator for optimization cycles, qasm simulator for final sampling of output, $R_y$ variational circuit of depth 2, penalty factor $10^3$, $\alpha = 25\%$, 2 optimization cycles, 50 function evaluations per cycle.]}
}
\label{fig:settlement-1-2-results}
\end{figure*}

The first problem instance, shown in Fig.~\ref{fig:settlement-1-2-diagrams}(a), does not feature a credit and collateral facility, while the second instance, shown in Fig.~\ref{fig:settlement-1-2-diagrams}(b), utilizes one for both accounts held by the first party. Starting with the first instance, we solve the following:
\begin{eqnarray} \label{eq:settlement-1-objective}
\max && (x_1 + x_2 + x_3) \\
&& \hspace{-3em} \textrm{subject to} \nonumber\\\nonumber
\textrm{$P_1$:} && (2, 0) + (-1,2) \, x_1 + (-1,2) \, x_3 \ge (0,0) \\\nonumber
\textrm{$P_2$:} && (0, 3) + (1,-2) \, x_1 + (1,-2) \, x_2 \ge (0,0) \\\nonumber
\textrm{$P_3$:} && (0, 0) + (-1,2) \, x_2 + (1,-2) \, x_3 \ge (0,0) \\\nonumber
\end{eqnarray}
where, for each party $P_i$, the account balances, transaction values, and account credit limits are indicated by vectors $(C,S)$ specifying the amounts of currency $C$ and security $S$. The optimal solution settles transactions $T_2, T_3$, that is, $(x_1, x_2, x_3) = 011$, while a greedy algorithm may be able to settle only a single transaction, $T_1$.

Next, we consider the batch of three transactions in the presence of a simplified credit and collateral facility. In this model, illustrated in Fig.~\ref{fig:settlement-1-2-diagrams}(b), the two accounts held by party $P_1$ can contribute to and draw from a shared credit pool $F_1$ containing a net zero balance. We adjust the balance of $P_1$ such that no transactions can settle without the credit and collateral facility, while the optimal solution settles $T_2$ and $T_3$ with the facility (note that settling only $T_1$ is a valid solution, though not optimal). The constraints for $P_1$ are then modified from those in Eq.~\eqref{eq:settlement-1-objective} to
\begin{eqnarray} \label{eq:settlement-2-objective}
\max && (x_1 + x_2 + x_3) \\
&& \hspace{-3em} \textrm{subject to} \nonumber\\\nonumber
\textrm{$P_1$:} && (0, 1) + (-1,2) \, x_1 + (-1,2) \, x_3 \ge (l_C, l_S) \\\nonumber
\textrm{$F_1$:} && \, l_C + 2 l_S = 0, ~~ l_C \ge -2, ~~ l_S \ge 0 \\\nonumber
\textrm{$P_2$:} && (0, 3) + (1,-2) \, x_1 + (1,-2) \, x_2 \ge (0,0) \\\nonumber
\textrm{$P_3$:} && (0, 0) + (-1,2) \, x_2 + (1,-2) \, x_3 \ge (0,0).\nonumber
\end{eqnarray}
Party $P_1$ now has an effective credit line for each account, $l_C$ and $l_S$.
We assume that the credit and collateral facility $F_1$ has a zero net balance so that $l_C$ and $l_S$ sum to zero with the appropriate exchange factors between accounts. 
In particular, the credit limit $l_C$ on the currency account cannot extend past $-1S \times (2 C/S) = -2C$, where $1S$ is the starting balance of the security account and $1 S = 2 C$ is the exchange factor. A similar analysis holds for the security account: $l_S \ge 0$.

The final problem instance, illustrated in Fig.~\ref{fig:settlement-3-diagrams}, involves a batch of seven payments of a single currency between six parties. The objective and full list of constraints is the following set of equations:
\begin{eqnarray} \label{eq:settlement-3-objective}
\max && (x_1 + x_2 + x_3 + x_4 + x_5 + x_6 + x_7) \\
&& \hspace{-3em} \textrm{subject to} \nonumber\\\nonumber
\textrm{$P_1$:} && 5 - 4 x_1 - 3 x_2 - 2 x_3 \ge 0 \\\nonumber
\textrm{$P_2$:} && 1 + 4 x_1 - 3 x_4 - 3 x_5 \ge 0 \\\nonumber
\textrm{$P_3$:} && 2 + 3 x_2 \ge 0 \\\nonumber
\textrm{$P_4$:} && 3 + 2 x_3 \ge 0 \\\nonumber
\textrm{$P_5$:} && 2 + 3 x_4 - 6 x_6 + 4 x_7 \ge 0 \\\nonumber
\textrm{$P_6$:} && 1 + 3 x_5 + 6 x_6 - 4 x_7 \ge 0,\nonumber
\end{eqnarray}
where the constraints for $P_3$ and $P_4$ are always satisfied and can be removed from the model. Several optimal solutions exist, for example with payments $T_2, T_3, T_6, T_7$ settling. In the next section, we discuss numerical results for each of these three problem instances.

%%%%%%%%%%%%%%%%%%%%%%%%%%%%%%%%%%%%%%%%%%%%%%%%%%%%%%%%%%%%%%%%%%%%%%%%%%%
\section{\label{sec:results} Results}
%%%%%%%%%%%%%%%%%%%%%%%%%%%%%%%%%%%%%%%%%%%%%%%%%%%%%%%%%%%%%%%%%%%%%%%%%%%

\begin{figure}[b]
\centering
\includegraphics[width=\linewidth]{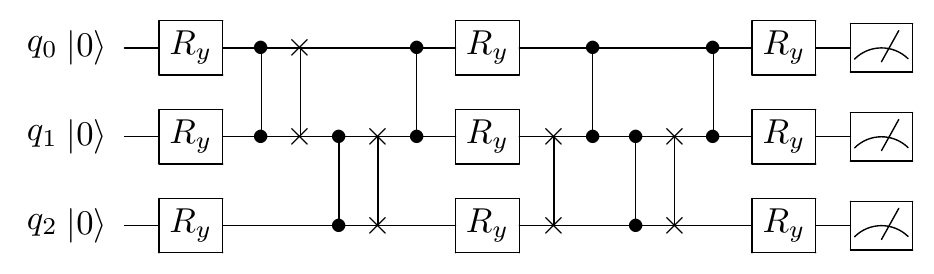}
\caption{Variational circuit used for the first two problem instances of three delivery-versus-payment transactions between three parties without and with a simplified credit and collateral facility, Eqs.~\eqref{eq:settlement-1-objective} and~\eqref{eq:settlement-2-objective}, respectively. The depth $d=2$ variational form is composed of $d+1$ layers of parametrized $R_y$ gates (parameters omitted in the figure) interleaved with $d$ fully-connected entangling blocks of controlled-phase gates, for a total of $n (d+1)$ independent angles, for $n$ qubits. The linear connectivity of the \emph{Valencia} quantum processor introduces {\scshape swap} gates for two-qubit gates between nonadjacent qubits. In practice, this circuit can be further optimized, e.g., cancelling a pair of {\scshape cnot}s from an adjacent {\scshape swap} and {\scshape cnot}. A similar circuit is used for simulation of the third problem instance of seven payments between six parties, but with depth $d=3$ and $n=7$ qubits.}
\label{fig:circuit}
\end{figure}

We solve each problem instance described in Sec.~\ref{sec:transaction_settlement} using the hybrid/quantum classical algorithms outlined in Sec.~\ref{sec:algorithms}. Following \cite{Nannicini2019}, we use COBYLA as the classical optimizer, since it is well suited to handle continuous variable constraints. 
For each problem instance, we classically simulate the hybrid quantum-classical algorithm using Qiskit \cite{Qiskit} and, for the first two problem instances, compare the simulations to runs on the \emph{Valencia} 5-qubit quantum processor of the IBM Quantum Computation Center. 
We employ readout-error mitigation using \emph{Qiskit Ignis} \cite{Dewes2012, Qiskit} to correct measurement errors.
The variational circuit we use, shown in Fig.~\ref{fig:circuit}, is composed of $d+1$ layers of single-qubit $Y$-rotations interleaved with $d$ blocks of controlled-$Z$ gates \cite{kandala2017hardware}. 
We set a quadratic penalty factor of $\lambda=10^3$ for the problem constraints [see Eq.~\eqref{eq:transaction_settlement_2}], and utilize the CVaR aggregation function mentioned in Sec.~\ref{sec:algorithms} with $\alpha = 25\%$ in place of the sample average for 8,192 samples drawn from each trial wavefunction. 

The results of the quantum algorithm for the first two problem instances using a simulator and the \emph{Valencia} quantum processor, are shown in Fig.~\ref{fig:settlement-1-2-results}. 
%We distinguish between feasible (solid green bars) and infeasible (hatched grey bars) candidate solutions, which indicate the transactions to settle, and sort left to right by decreasing total number of transactions settled.

\begin{figure*}[t]
\centering
\includegraphics[width=0.85\linewidth]{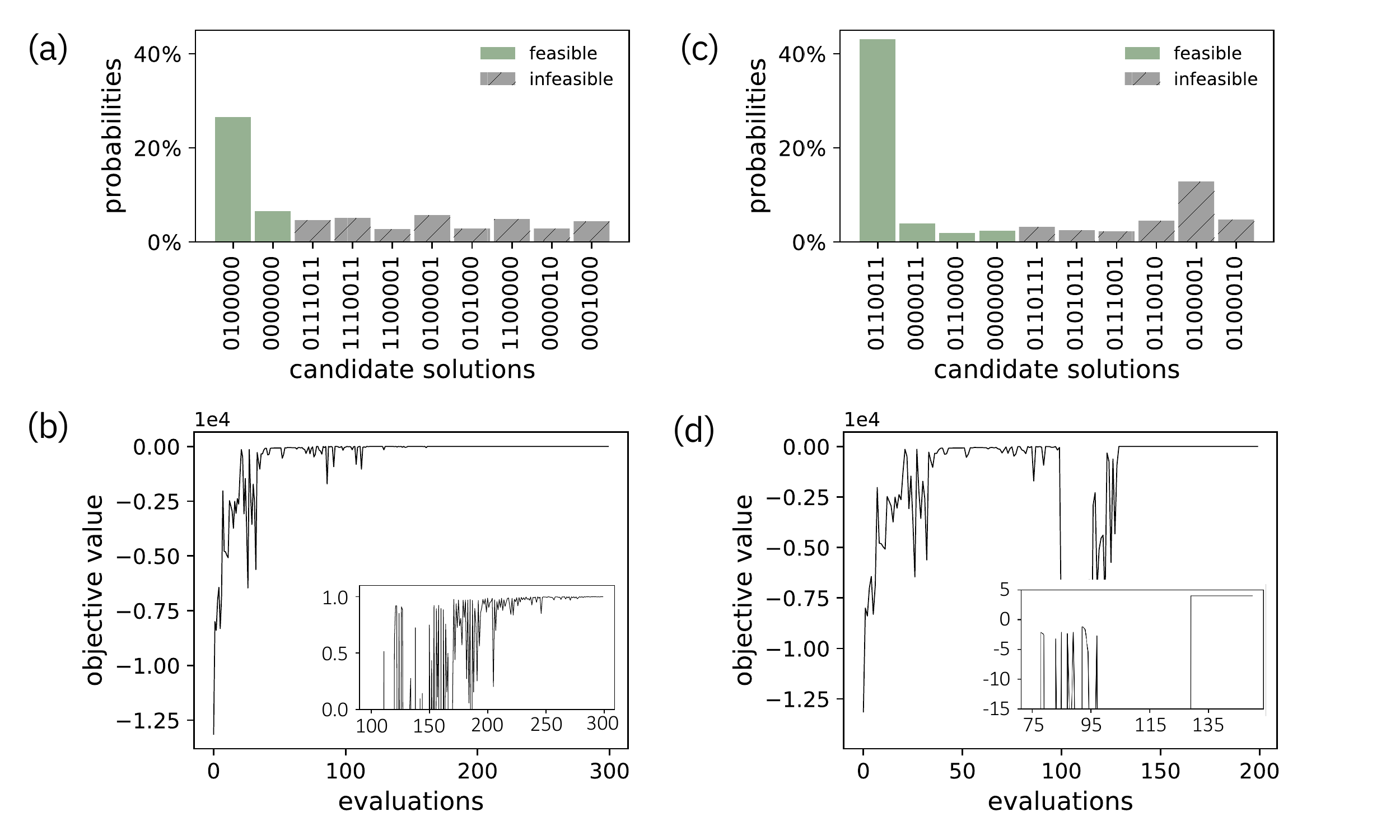}
\caption{
Results for six parties and seven payments of a single currency using classical simulation. The problem is first approached without (a, b) and then with (c, d) explicit handling of slack variables, i.e. the heuristic described in Sec.~\ref{sec:algorithms}. 
In the first approach, we apply COBYLA with 300 iterations, while in the second, we run 1 cycle with 100 + 100 iterations of COBYLA (the first 100 iterations on the full problem and the second 100 iterations on the parameters of the variational form only).
We show here the probabilities of the ten most frequent states resulting from the optimization process (a, c), where candidate solutions are grouped into feasible (green bars) and infeasible (grey hatched bars) sets and then sorted left to right by descending value of the objective function (number of trades settled).
In (a), the feasible solutions sampled are sub-optimal, the best one having a single payment settled.
In (b), an optimal solution with 4 payments settled is found with a probability larger than 40\%.
In addition, we show the progress during the optimization, where it can clearly be seen how the explicit handling of slack variables (d) achieves the optimal solution while the direct application of COBYLA gets stuck in a local optimum (b).
Results shown here used a $R_y$ variational circuit of depth $d=3$, a penalty factor $\lambda=10^3$ for adding constraints to the objective function, and the CVaR aggregation function with $\alpha = 25\%$.}
\label{fig:settlement-3-results}
\end{figure*}

We compare the results for the first problem \eqref{eq:settlement-1-objective} from classical simulation in Fig.~\ref{fig:settlement-1-2-results}(a) to those found from running on \emph{Valencia} in Fig.~\ref{fig:settlement-1-2-results}(b), and similarly for the second problem \eqref{eq:settlement-2-objective} in Figs.~\ref{fig:settlement-1-2-results}(c) and~\ref{fig:settlement-1-2-results}(d), respectively. For both problem instances, with and without the credit and collateral facility, we find good agreement between simulation and experiment. 
The optimal solution that settles $T_2$ and $T_3$ (the bitstring $011$) appears with the highest probability.

For the third problem instance \eqref{eq:settlement-3-objective}, we simulate classically and compare the algorithm with explicit handling of slack variables and without, as described in the heuristic of Sec.~\ref{sec:algorithms}. In the latter case, we find that the objective function never reaches the maximum value of four (i.e. four settled transactions), but saturates at unity as seen in Fig.~\ref{fig:settlement-3-results}(b). As a result, only sub-optimal feasible solutions are sampled, see Fig.~\ref{fig:settlement-3-results}(a). In contrast, with explicit handling of slack variables, Fig.~\ref{fig:settlement-3-results}(d) shows how the objective function achieves the maximum value within the second part (step 3, evaluations 100-200) of the first cycle, and that we now sample a globally optimal solution settling payments $T_2, T_3, T_6, T_7$ (the bitstring 0110011) in~\ref{fig:settlement-3-results}(c). 
Note that the heuristic finds only one out of the three possible globally optimal solutions.
To influence the actual solution found, we could adjust the weights $w_i$ in Eq.~\eqref{eq:transaction_settlement_1}, for instance, to prioritize transactions with a larger volume.

All the algorithms described here have been implemented in Qiskit \cite{Qiskit}, which has been used to access the classical simulator as well as the quantum hardware.

%%%%%%%%%%%%%%%%%%%%%%%%%%%%%%%%%%%%%%%%%%%%%%%%%%%%%%%%%%%%%%%%%%%%%%%%%%%
\section{\label{sec:conclusions} Conclusions}
%%%%%%%%%%%%%%%%%%%%%%%%%%%%%%%%%%%%%%%%%%%%%%%%%%%%%%%%%%%%%%%%%%%%%%%%%%%

In this paper we extended the existing work on quantum optimization to the class of Mixed Binary Optimization and propose the first hybrid quantum/classical heuristics to address such problems.
This allows us, for instance, to model inequality constraints, which significantly extends the applicability of these algorithms.
Inequality constraints are crucial for many applications such as portfolio optimization, where to date, quantum algorithms were always assuming a cardinality equality constraint instead of a real budget constraint \cite{Venturelli2018a, Barkoutsos2019}.
We tested our algorithms on the Transaction Settlement problem for batches of payment and delivery-versus-payment securities transactions, which is a difficult and important optimization challenge in capital markets.
Better algorithms could increase settlement efficiency (in terms of the number of transactions settled for a given batch), thereby minimizing the time intervals between trade and settlement. This could reduce replacement cost risk (the risk of loss of unrealized gains due to delay in settlement), reduce liquidity risk (substantial liquidity pressures can emerge if a participant fails to settle its net funds debit positions), and reduce credit risk (particularly if there is a decline in value of the securities).

The algorithms presented in this paper are exploiting only very little of the structure of a problem.
Although this allows for great flexibility, it also clearly indicates directions of future research.
By restricting to problems with certain properties such as, for instance, convexity in the continuous variables, it may be possible to derive algorithms that outperform the current state-of-the-art and maybe even derive some guarantees on convergence under assumptions of the underlying quantum algorithms.
The proposed structure also allows for the use of problem specific variational forms, such as those proposed by QAOA, which may further improve the performance.

%%%%%%%%%%%%%%%%%%%%%%%%%%%%%%%%%%%%%%%%%%%%%%%%%%%%%%%%%%%%%%%%%%%%%%%%%%%
%\begin{acknowledgments}
%\end{acknowledgments}
%%%%%%%%%%%%%%%%%%%%%%%%%%%%%%%%%%%%%%%%%%%%%%%%%%%%%%%%%%%%%%%%%%%%%%%%%%%

%\subsection*{Author contributions}
%All authors researched, collated, and wrote this paper.

%\subsection*{Competing interests}
%The authors declare no competing interest.

%%%%%%%%%%%%%%%%%%%%%%%%%%%%%%%%%%%%%%%%%%%%%%%%%%%%%%%%%%%%%%%%%%%%%%%%%%%
%\appendix
%%%%%%%%%%%%%%%%%%%%%%%%%%%%%%%%%%%%%%%%%%%%%%%%%%%%%%%%%%%%%%%%%%%%%%%%%%%

%%%%%%%%%%%%%%%%%%%%%%%%%%%%%%%%%%%%%%%%%%%%%%%%%%%%%%%%%%%%%%%%%%%%%%%%%%%
\bibliography{bibliography}% Produces the bibliography via BibTeX.
%%%%%%%%%%%%%%%%%%%%%%%%%%%%%%%%%%%%%%%%%%%%%%%%%%%%%%%%%%%%%%%%%%%%%%%%%%%

\end{document}